\documentclass[aps,prl,preprint,tightenlines,superscriptaddress,showpacs,byrevtex]{revtex4}
\usepackage{graphicx}
\usepackage{ulem}

\def\figonescale{0.4}
\def\figtwoscale{0.25}

\makeatletter
\def\myep@[#1]#2{\resizebox{#1\textwidth}{!}{\includegraphics{#2}}}
\def\myeps{\@ifnextchar[{\myep@}{\myep@[1]}}
\makeatother

\def\NBBused{657\EP6}
\def\NBBusedwitherror{(657\pm9)\EP6}

\def\drcut{0.5\cm}      
\def\dzcut{3.0\cm}      
\def\ptrkcut{100\MeVc}  
         
\def\EgampiZmin{50\MeV} 
\def\MpiZwindow{16}

\def\MKpiZmin{0.90}

\def\coshelRP{0.80}    
\def\coshelRZ{0.75}    
\def\coshelOM{0.83}    

\def\Egammin{1.8}      
\def\Egammax{3.4}

\def\Mrhomin{640}    
\def\Mrhomax{890}    
\def\Momegamin{760}  
\def\Momegamax{800}  
\def\MKstarmin{820}  
\def\MKstarmax{970}  

\def\EffKaon{85\%}
\def\EffPionRP{86\%}
\def\EffPionRZ{87\%}
\def\EffPionOM{89\%}

\def\FakePionRP{8.3\%}
\def\FakePionRZ{8.5\%}

\def\BrBtoKPGstat{384 \pm 17}
\def\BrBtoKZGstat{378 \pm 8} 

\def\ratioBtoRZGtoKZG{0.0206\PM{0.0045}{0.0043}\PM{0.0014}{0.0016}}
\def\ratioBtoRGtoKG{0.0302\PM{0.0060}{0.0055}\PM{0.0026}{0.0028}}
\def\ratioBtoROGtoKG{0.0284\pm0.0050\PM{0.0027}{0.0029}}

\def\NSBtoRPG{45.8 \PM{15.2}{14.5} \PM{2.6}{3.9}}
\def\BrBtoRPG{8.7 \PM{2.9}{2.7} \PM{0.9}{1.1}} 

\def\NSBtoRZG{75.7 \PM{16.8}{16.0} \PM{5.1}{6.1}}
\def\BrBtoRZG{7.8 \PM{1.7}{1.6} \PM{0.9}{1.0}} 

\def\NSBtoOMG{17.5 \PM{8.2}{7.4} \PM{1.1}{1.0}}
\def\BrBtoOMG{4.0 \PM{1.9}{1.7}\pm 1.3} 

\def\BrBtoRAG{12.1\PM{2.4}{2.2}\pm 1.2} 
\def\BrBtoROG{11.4\pm 2.0 \PM{1.0}{1.2} }

\def\signifRAconv{5.8}
\def\signifROconv{6.2}

\def\AcpBtoRPG{-0.11\pm{0.32}\pm{0.09}}
\def\AIBtoRG{-0.48\PM{0.21}{0.19}\PM{0.08}{0.09}}

\def\VtdVtsrw{0.195 \PM{0.020}{0.019} \expl \pm 0.015 \theo}

\def\EffRP{8.03\pm0.59}
\def\EffRZ{14.81\pm0.95}
\def\EffOM{6.58\pm0.76}

\def\EffLRsigRP{35\%}
\def\EffLRsigRZ{51\%}
\def\EffLRsigOM{43\%}

\def\signifRPconv{3.3}
\def\signifRZconv{5.0}
\def\signifOMconv{2.6}

\def\KP{K^+}
\def\KM{K^-}
\def\KS{{K^0_S}}
\def\KL{{K^0_L}}

\def\piZ{{\pi^0}}
\def\piP{{\pi^+}}
\def\piM{{\pi^-}}
\def\rhoP{{\rho^+}}
\def\rhoZ{{\rho^0}}
\def\Kstar{{K^*}}
\def\KstarP{{K^{*+}}}
\def\KstarZ{{K^{*0}}}
\def\omegaG{{\omega\gamma}}
\def\rhoG{{\rho\gamma}}
\def\omegG{{\omega\gamma}}
\def\rhoPG{{\rhoP\gamma}}
\def\rhoZG{{\rhoZ\gamma}}
\def\ROG{{(\rho,\omega)\gamma}}
\def\KstarG{{\Kstar\gamma}}
\def\KstarZG{{\KstarZ\gamma}}
\def\KstarPG{{\KstarP\gamma}}
\def\qqbar{q\overline{q}{}}

\def\epem{e^+e^-{}}

\def\btodgamma{b\to d\gamma}

\def\rhoM{\rho^-}
\def\rhoP{\rho^+}
\def\rhoZ{\rho^0}
\def\rhoG{\rho\gamma}
\def\rhoMG{\rhoM\gamma}
\def\rhoPG{\rhoP\gamma}
\def\rhoZG{\rhoZ\gamma}

\def\BtoRG{B\to \rho\gamma}
\def\BtoRPG{B^+\to \rho^+\gamma}

\def\BtoRZG{B^0\to \rho^0\gamma}
\def\BtoROG{B\to (\rho,\omega)\gamma}
\def\BtoOMG{B^0\to \omega\gamma}
\def\BtoOG{B\to \omega\gamma}

\def\BtoKG{B\to K^*\gamma}
\def\BtoKZG{B^0\to K^{*0}\gamma}
\def\BtoKPG{B^+\to K^{*+}\gamma}

\def\BtoXsgamma{B\to X_s\gamma}

\def\btoc{b\to c}

\def\cm{\mbox{~cm}}

\def\GeV{\mbox{~GeV}}

\def\GeVcc{\mbox{~GeV}/c^2}
\def\MeV{\mbox{~MeV}}
\def\MeVc{\mbox{~MeV}/c}
\def\MeVcc{\mbox{~MeV}/c^2}

\def\Vtd{V_{td}}
\def\Vts{V_{ts}}

\def\Br{{\cal B}}
\def\Acp{A_{CP}}

\def\Lpi{{\cal L}_\pi}
\def\LK{{\cal L}_K}

\def\Mbc{M_{\rm bc}}
\def\DeltaE{\Delta{E}}
\def\Ebeam{E^*_{\rm beam}{}}

\def\pvecB{\vec{p}{}^{\;*}_B}

\def\MKpi{M_{K\pi}}

\def\Egamma{E^*_\gamma}

\def\tauBratio{{\tau_{B^+}\over\tau_{B^0}}}
\def\tauBratiorev{{\tau_{B^0}\over2\tau_{B^+}}}

\def\piZeta{\piZ/\eta}

\def\thetaB{{\theta^*_B}}
\def\cosB{{\cos\thetaB}}

\def\calF{{\cal F}}
\def\calLs{{\cal L}_s}
\def\calLc{{\cal L}_c}

\def\Momega{M_{\piP\piM\piZ}}

\def\calR{{\cal R}}

\def\thetahel{\theta_{\rm hel}}
\def\coshel{\cos\thetahel}

\def\Lzero{{\cal L}_0}
\def\Lmax{{\cal L}_{\rm max}}

\def\PM#1#2{\,^{+#1}_{-#2}{}}
\def\EM#1{\times10^{-#1}}
\def\EP#1{\times10^{#1}}

\def\theo{\rm(th.)}
\def\expl{\rm(exp.)}

\def\etal{\textit{et al.}}
\def\Journal#1#2#3#4{{#1} {\bf #2}, #3 (#4)}
\def\NIMA{Nucl. Instrum. Meth. A}
\def\NPB{Nucl. Phys. B}
\def\PLB{Phys. Lett. B}
\def\PRL{Phys. Rev. Lett.}
\def\PRD{Phys. Rev. D}
\def\ZPC{Z. Phys. C}
\def\EPJC{Eur. Phys. J. C}
\def\JPG{J. Phys. G}

\begin{document}

\preprint{\vbox{ \hbox{   }
                        \hbox{Belle Preprint 2008-13}
                                                \hbox{KEK Preprint
                        2008-07}
                        }}

\title{
  \vspace*{5mm}Measurement of branching fractions, isospin and
  $CP$-violating asymmetries for exclusive \boldmath$b \to d\gamma$
   modes
}

\begin{abstract}
  We report new measurements of the decays $\BtoRPG$, $\BtoRZG$ and
  $\BtoOMG$ using a data sample of $\NBBused$ $B$ meson pairs accumulated
  with the Belle detector at the KEKB $\epem$ collider.  We measure
  branching fractions $\Br(\BtoRPG)= (\BrBtoRPG)\EM7$, $\Br(\BtoRZG)=
  (\BrBtoRZG)\EM7$ and $\Br(\BtoOMG)= (\BrBtoOMG)\EM7$. We also
 report the  isospin
  asymmetry $\Delta(\rho\gamma)=\AIBtoRG$; and the first measurement of
 the direct $CP$-violating asymmetry
  $\Acp(\BtoRPG)=\AcpBtoRPG$, where the first and second errors are
  statistical and systematic, respectively.
\end{abstract}


\affiliation{Budker Institute of Nuclear Physics, Novosibirsk}
\affiliation{Chiba University, Chiba}
\affiliation{University of Cincinnati, Cincinnati, Ohio 45221}
\affiliation{Department of Physics, Fu Jen Catholic University, Taipei}
\affiliation{Justus-Liebig-Universit\"at Gie\ss{}en, Gie\ss{}en}
\affiliation{The Graduate University for Advanced Studies, Hayama}
\affiliation{Hanyang University, Seoul}
\affiliation{University of Hawaii, Honolulu, Hawaii 96822}
\affiliation{High Energy Accelerator Research Organization (KEK), Tsukuba}
\affiliation{Hiroshima Institute of Technology, Hiroshima}
\affiliation{University of Illinois at Urbana-Champaign, Urbana, Illinois 61801}
\affiliation{Institute of High Energy Physics, Chinese Academy of Sciences, Beijing}
\affiliation{Institute of High Energy Physics, Vienna}
\affiliation{Institute of High Energy Physics, Protvino}
\affiliation{Institute for Theoretical and Experimental Physics, Moscow}
\affiliation{J. Stefan Institute, Ljubljana}
\affiliation{Kanagawa University, Yokohama}
\affiliation{Korea University, Seoul}
\affiliation{Kyoto University, Kyoto}
\affiliation{Kyungpook National University, Taegu}
\affiliation{\'Ecole Polytechnique F\'ed\'erale de Lausanne (EPFL), Lausanne}
\affiliation{Faculty of Mathematics and Physics, University of Ljubljana, Ljubljana}
\affiliation{University of Maribor, Maribor}
\affiliation{University of Melbourne, School of Physics, Victoria 3010}
\affiliation{Nagoya University, Nagoya}
\affiliation{Nara Women's University, Nara}
\affiliation{National Central University, Chung-li}
\affiliation{National United University, Miao Li}
\affiliation{Department of Physics, National Taiwan University, Taipei}
\affiliation{H. Niewodniczanski Institute of Nuclear Physics, Krakow}
\affiliation{Nippon Dental University, Niigata}
\affiliation{Niigata University, Niigata}
\affiliation{University of Nova Gorica, Nova Gorica}
\affiliation{Osaka City University, Osaka}
\affiliation{Osaka University, Osaka}
\affiliation{Panjab University, Chandigarh}
\affiliation{RIKEN BNL Research Center, Upton, New York 11973}
\affiliation{Saga University, Saga}
\affiliation{University of Science and Technology of China, Hefei}
\affiliation{Seoul National University, Seoul}
\affiliation{Sungkyunkwan University, Suwon}
\affiliation{University of Sydney, Sydney, New South Wales}
\affiliation{Tata Institute of Fundamental Research, Mumbai}
\affiliation{Toho University, Funabashi}
\affiliation{Tohoku Gakuin University, Tagajo}
\affiliation{Department of Physics, University of Tokyo, Tokyo}
\affiliation{Tokyo Institute of Technology, Tokyo}
\affiliation{Tokyo Metropolitan University, Tokyo}
\affiliation{Tokyo University of Agriculture and Technology, Tokyo}
\affiliation{Virginia Polytechnic Institute and State University, Blacksburg, Virginia 24061}
\affiliation{Yonsei University, Seoul}
\author{N.~Taniguchi}\affiliation{Kyoto University, Kyoto} 
\author{M.~Nakao}\affiliation{High Energy Accelerator Research Organization (KEK), Tsukuba} 
\author{S.~Nishida}\affiliation{High Energy Accelerator Research Organization (KEK), Tsukuba} 
   \author{I.~Adachi}\affiliation{High Energy Accelerator Research Organization (KEK), Tsukuba} 
   \author{H.~Aihara}\affiliation{Department of Physics, University of Tokyo, Tokyo} 
   \author{K.~Arinstein}\affiliation{Budker Institute of Nuclear Physics, Novosibirsk} 
   \author{T.~Aushev}\affiliation{\'Ecole Polytechnique F\'ed\'erale de Lausanne (EPFL), Lausanne}\affiliation{Institute for Theoretical and Experimental Physics, Moscow} 
   \author{T.~Aziz}\affiliation{Tata Institute of Fundamental Research, Mumbai} 
   \author{A.~M.~Bakich}\affiliation{University of Sydney, Sydney, New South Wales} 
   \author{V.~Balagura}\affiliation{Institute for Theoretical and Experimental Physics, Moscow} 
% ...
\collaboration{The Belle Collaboration}

\maketitle
\begin{figure}[ht]
\begin{center}
\myeps[\figonescale]{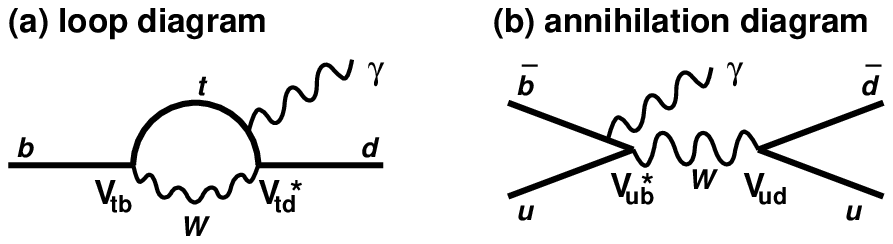}
\caption{(a) Loop diagram for $\btodgamma$ and (b) annihilation diagram,
  which contributes only to $\BtoRPG$.}
\label{fig:diagram}
\end{center}
\end{figure}

The $\btodgamma$ process, which proceeds via a loop diagram
(Fig.~\ref{fig:diagram}(a)) in the Standard Model (SM), provides a
valuable tool to search for physics beyond the SM, since the
loop diagram may also involve virtual heavy non-SM
particles~\cite{bib:rhogam-bsm}.  The process has been observed in the
exclusive modes $\BtoRG$ and $\BtoOG$ by Belle~\cite{bib:belle-rhogam}
and Babar~\cite{bib:babar-rhogam}. 
 Branching fractions for these modes have been used to constrain the ratio of
Cabibbo-Kobayashi-Maskawa (CKM) matrix elements~\cite{bib:ckm}
$|\Vtd/\Vts|$; 
 a non-SM effect may be
observed as a deviation of $|\Vtd/\Vts|$ from
 the expectation based on measurements of other CKM matrix elements
and unitarity of the matrix~\cite{bib:ckmfitter}.
An additional contribution from an annihilation diagram
(Fig.~\ref{fig:diagram}(b)) may induce a direct $CP$-violating asymmetry
in $\BtoRPG$,
and an isospin asymmetry between $\BtoRG$ modes; the latter can be used to
constrain the CKM unitarity triangle angle $\phi_3$~\cite{bib:pball}.
These quantities are also sensitive to physics beyond the SM~\cite{bib:alilungi}.
In this paper, we report new measurements of the $\BtoRG$ and
$\BtoOG$ processes using a data sample of $\NBBusedwitherror$ $B$ meson pairs
accumulated at the $\Upsilon(4S)$ resonance.  With a data sample
 almost twice as large
and an improved analysis procedure, these results supersede those in~\cite{bib:belle-rhogam}.

The data are obtained in $\epem$ annihilation at the KEKB
energy-asymmetric (3.5 on 8 GeV) collider~\cite{bib:kekb} and collected
with the Belle detector~\cite{bib:belle-detector}.  The Belle detector
includes a silicon vertex detector (SVD), a central drift chamber (CDC),
aerogel threshold Cherenkov counters (ACC), time-of-flight (TOF)
scintillation counters, and an electromagnetic calorimeter (ECL)
comprised of
CsI(Tl) crystals located inside a 1.5 T superconducting solenoid coil.
An iron flux-return located outside of the coil is instrumented to
identify $\KL$ and muons (KLM).

We reconstruct three signal modes, $\BtoRPG$, $\BtoRZG$ and $\BtoOMG$,
and two control samples, $\BtoKPG$ and $\BtoKZG$.  Charge-conjugate
modes are implicitly included unless otherwise stated.  The following
decay modes are used to reconstruct the intermediate states:
$\rhoP\to\piP\piZ$, $\rhoZ\to\piP\piM$, $\omega\to\piP\piM\piZ$,
$\KstarP\to\KP\piZ$, $\KstarZ\to\KP\piM$, and $\piZ\to\gamma\gamma$.

Photon candidates are reconstructed from ECL energy clusters having a
photon-like shape and no associated charged track.  A photon with an
 $\Upsilon(4S)$ 
center-of-mass (c.m.)\ energy $(\Egamma)$ in the range
$[\Egammin,\Egammax]\GeV$ is selected as the primary photon candidate.
Photons detected by the endcap ECL, which were excluded in the previous
analysis, are also used.  To suppress backgrounds from
$\piZ/\eta\to\gamma\gamma$ decays, we apply a veto algorithm based on
the likelihood calculated for every photon pair consisting of the primary photon
and another photon.
 We also reject the primary photon candidate if the ratio of the
energy in the $3\times3$ crystal array, centered on the crystal with the
maximum energy, to that in the $5\times5$ array is less than 0.95. 

Neutral pions are formed from photon pairs with invariant masses within
$\pm \MpiZwindow\MeVcc$ (${\sim}3\sigma$) of the $\piZ$ mass.  We
require the energy of each photon to be greater than $\EgampiZmin$, and
the cosine of the angle between the two photons in the laboratory frame
to be greater than 0.58 (0.40) for the $\piZ$ from $\rhoP$ ($\omega$).
The photon momenta are then recalculated with a $\piZ$ mass constraint.

Charged pions and kaons are selected from tracks in the CDC and SVD.
Each track is required to have a transverse momentum greater than
$\ptrkcut$ and a distance of closest approach to the interaction point
within $\drcut$ in radius and  within $\pm\dzcut$ along the positron beam ($z$-)
axis.
{We use a likelihood ratio $\Lpi/(\Lpi + \LK) < 0.3$ for pions and
$> 0.6$ for kaons, where the pion and kaon likelihoods $\Lpi$ and $\LK$ are
determined from ACC, TOF and CDC information.}
The criteria have efficiencies of $\EffPionRP$, $\EffPionRZ$ and
$\EffPionOM$ for a pion from $\rhoP$, $\rhoZ$ and $\omega$,
respectively; the misidentification probability for a kaon is $\FakePionRP$
($\FakePionRZ$) for $\rhoP$ ($\rhoZ$).
Kaons for $\Kstar$ candidates are selected with an efficiency of $\EffKaon$.
Invariant masses for the $\rho$, $\omega$ and $\Kstar$ candidates are
required to be within windows of $[\Mrhomin, \Mrhomax]\MeVcc$,
$[\Momegamin, \Momegamax]\MeVcc$, and $[\MKstarmin, \MKstarmax]\MeVcc$,
respectively.

Candidate $B$ mesons are reconstructed by combining a $\rho$ or $\omega$
candidate with the primary photon and calculating two variables: the
beam-energy constrained mass $\Mbc = \sqrt{ (\Ebeam/c^2)^2 -
|\pvecB/c|^{2}}$, and the energy difference $\Delta E = E^*_{B} -
\Ebeam$.  Here, $\pvecB$ and $E^*_B$ are the {c.m.}\ momentum and energy of
the $B$ candidate, and $\Ebeam$ is the {c.m.}\ beam energy.  To improve
resolution, the magnitude of the photon momentum is replaced by $(\Ebeam
- E_{\rho/\omega}^*)/c$ when the momentum $\pvecB$ is calculated.

To optimize the event selection, we study Monte Carlo (MC) events in a
signal box defined as $5.273\GeVcc<\Mbc<5.285\GeVcc$ and
$|\DeltaE|<0.1\GeV$.  For each signal mode, we choose selection criteria to maximize
$N_S/\sqrt{N_S+N_B}$, where $N_S$ and $N_B$
are the expected signal and the sum of the background yields.

The dominant background arises from continuum events
($\epem\to\qqbar(\gamma)$, $q=u,d,s,c$), where a random combination of a
$\rho$ or $\omega$ candidate with a photon forms a $B$ candidate.  We
suppress this using a Fisher discriminant ($\calF$) calculated from
modified Fox-Wolfram moments~\cite{bib:ksfw} and other variables, i.e., the cosine of the
polar angle ($\cosB$) of the $B$ direction, the distance along the
$z$-axis ($\Delta z$) between the signal vertex and that of the rest of
the event 
and, in addition, $\Momega$ and Dalitz plot variables
for the $\omegaG$ mode.
For each of these quantities, we construct likelihood distributions for
signal and continuum events.  The distributions are determined from MC
samples.

 From these likelihood distributions we form likelihoods
$\calLs$ and $\calLc$ for signal and continuum background,
respectively. 
In addition, we use a flavor-tagging quality variable $r$ that
indicates the level of confidence in the $B$-flavor determination as
described in Ref.~\cite{bib:hamlet}.  In the $(r,\calR)$ plane defined
by the tagging quality $r$ and the likelihood ratio $\calR={\cal
L}_s/({\cal L}_s+{\cal L}_c)$, signal tends to populate the edges at
$r=1$ and $\calR=1$, while continuum preferentially populates the edges
at $r=0$ and $\calR=0$.  We divide the events into six bins of $r$ (two
bins between 0 and 0.5, and four bins between 0.5 and 1) and determine the
minimum $\calR$ requirement for each bin.  In the $\rhoPG$ mode, if the
tagging-side flavor is the same as that of the signal side, we assign
the events to the lowest bin $0\le r<0.25$.
The $\calR$ criteria reject 98$\%$ of continuum background 
while retaining $\EffLRsigRP$, $\EffLRsigRZ$ and $\EffLRsigOM$ of the
$\rhoZG$, $\rhoPG$ and $\omegaG$ signals, respectively.
For the $\KstarPG$ ($\KstarZG$) mode, we use the criteria for the
$\rhoPG$ ($\rhoZG$) mode.

We consider the following backgrounds from $B$ decays: $\BtoKG$, other
$B\to X_s\gamma$ processes, decays with a $\piZeta$ ($B\to\rho\piZ$,
$\omega\piZ$, $\rho\eta$ and $\omega\eta$), other charmless hadronic $B$
decays, and $\btoc$ decay modes.  The $\BtoKG$ background can mimic the
$\BtoRG$ signal if the kaon from the $K^*$ is misidentified as a pion.
To suppress $\BtoKG$ events we calculate $\MKpi$, where the kaon mass is
assigned to the charged pion candidate; for $\rhoZG$,
the lower of the
two $\MKpi$ values is taken (misassignment tends to give a
higher $\MKpi$).
For the $\rhoPG$ mode, we reject the candidate if
$\MKpi<\MKpiZmin\GeVcc$, while for the $\rhoZG$
mode we use $\MKpi$ in the fit procedure to extract the signal
(note: $\MKpi>0.92\GeVcc$ is required when optimizing selection criteria).
 The $b \to s \gamma$ modes ($\BtoKG$ and other $\BtoXsgamma$ decays) contribute to the background
when the $\rho$ and $\omega$ candidates are formed from random
combinations of particles.  Decays with a $\piZeta$ can mimic the signal
if one of its daughter photons is not detected.  To suppress this
background, we reject the candidate if $|\coshel|>\coshelRP$,
$\coshelRZ$ and $\coshelOM$ for the $\rhoPG$, $\rhoZG$ and $\omegG$
mode, respectively, where the helicity angle $\thetahel$ is the angle
between the $\piP$ track (the normal to the $\omega$ decay plane) and the
$B$ momentum vector in the $\rho$ ($\omega$) rest frame.
 We study large MC samples and find no other distinctive hadronic $B$ decay background
sources. 

The reconstruction efficiency for each mode is defined as the fraction
of the signal remaining after all selection criteria are applied, where
the signal yield is determined from a fit to the sum of the signal and
continuum MC samples using the procedure described below.
We take the pion identification efficiency from a data sample of
$D^{*+}\to D^0\piP$, $D^0\to\KM\piP$.
  The total efficiencies are listed in Table~\ref{tbl:results}.
The systematic error in the efficiency is the quadratic sum of the
following contributions, estimated using control samples: the
uncertainty in the photon detection efficiency (2.4\%) as measured in
radiative Bhabha events; charged tracking efficiency (1.0\% per track)
from partially reconstructed $D^{*+}\to D^0\piP$, $D^0\to\KS\piP\piM$,
$\KS\to\piP(\piM)$; charged pion and kaon identification (0.5 to 0.6\%
per track) from $D^{*+}\to D^0\piP$, $D^0\to\KM\piP$; neutral pion
detection (4.6\%) from $\eta$ decays to $\gamma\gamma$, $\piP\piM\piZ$
and $3\piZ$; and $\calR$-$r$ and $\piZeta$ veto requirements (2.0 to
8.4\%) from $B\to D\piM$ with $D^0\to\KM\piP$, $\KS\omega$ and
$D^+\to\KM\piP\piP$.

We perform an unbinned extended maximum likelihood fit to $\Mbc$ and
$\DeltaE$ (and $\MKpi$ for the $\rhoZG$ mode) for candidates satisfying
$|\DeltaE|<0.5\GeV$ and $\Mbc>5.2\GeVcc$.  The fit is performed
individually for the three $\rhoG/\omegaG$ signal modes and the two
$\KstarG$ modes.  We describe the events in the fit region using the sum
of probability density functions for the signal, continuum, $\KstarG$ (for the $\rhoG$ modes
only), and other background hypotheses.  We use the distributions of MC
events in histograms to model the $\Mbc$-$\DeltaE$ shapes of $B$ decay
background components and the $\MKpi$ shapes for all components.

The signal distribution for the $\rhoZG$ and $\KstarZG$ modes is modeled as the product
of a Crystal Ball lineshape~\cite{bib:cbls} in $\DeltaE$ to reproduce
the asymmetric ECL energy response, a Gaussian in $\Mbc$, and an MC
histogram distribution for $\MKpi$.  For the $\rhoPG$, $\KstarPG$ and $\omegaG$
modes, we use the product of a Crystal Ball lineshape for $\DeltaE$ and
another Crystal Ball lineshape for $\Mbc$.  The signal parameters of
$\Mbc$ and $\DeltaE$ shapes for $\KstarG$ modes are determined from
fitting the data; for the $\rhoG/\omegaG$ modes, they are taken
from MC and calibrated
using the data/MC difference of the fits to the $\KstarPG$ and
$\KstarZG$ samples for the modes with and without a neutral pion,
respectively.

The continuum background component is modeled as the product of a linear
function in $\DeltaE$, an ARGUS function~\cite{bib:argus-function} in
$\Mbc$, and, for $\rhoZG$, an MC histogram for $\MKpi$.  The continuum
shape parameters and normalizations are mode dependent and allowed to
float.

There is significant $\KstarZG$ background in the $\rhoZG$ sample. 
This background is modeled
by the product of a two-dimensional $\Mbc$-$\DeltaE$ histogram and
an $\MKpi$ histogram.  Similarly, the $\KstarPG$ background for $\rhoPG$
is modeled by a two-dimensional $\Mbc$-$\DeltaE$ histogram.
In both cases, the $\DeltaE$ peak position is shifted from the
$\rho\gamma$ signal peak; this offset is determined from 
fitting the MC histogram shape to a $\KstarG$ data sample in which the pion
mass is assigned to kaons.
 The same $\KstarG$ sample together with the known kaon to pion
misidentification probability is also used to determine the size of the
$\KstarG$ background.

 Other $B$ decays are considered as an additional
background component when we extract the signal yield.  The levels of
these backgrounds are fixed using known branching fractions or upper
limits~\cite{bib:hfag2006}.

\begin{figure}[t]
    \begin{center}
       \myeps[\figtwoscale]{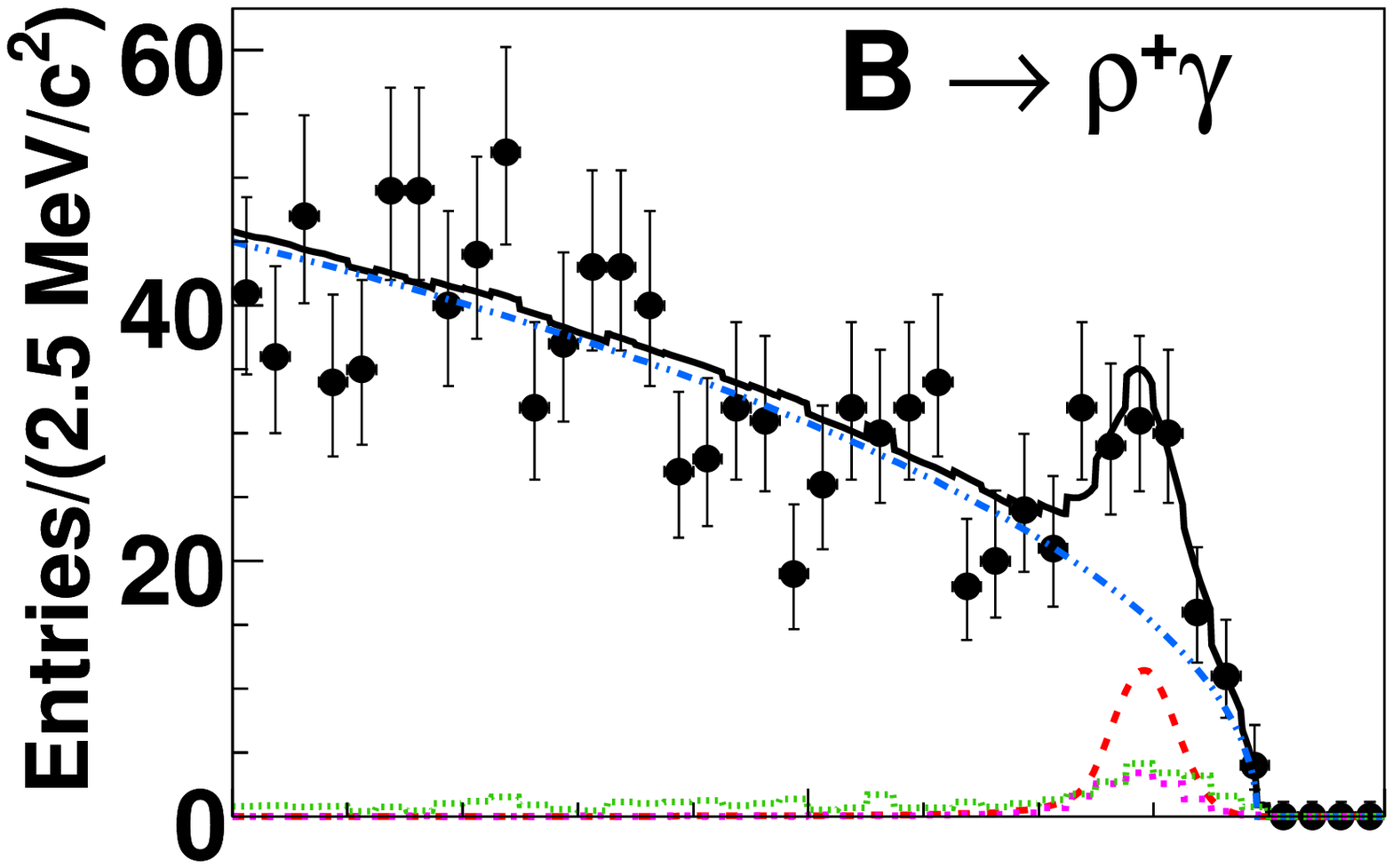}%
       \myeps[\figtwoscale]{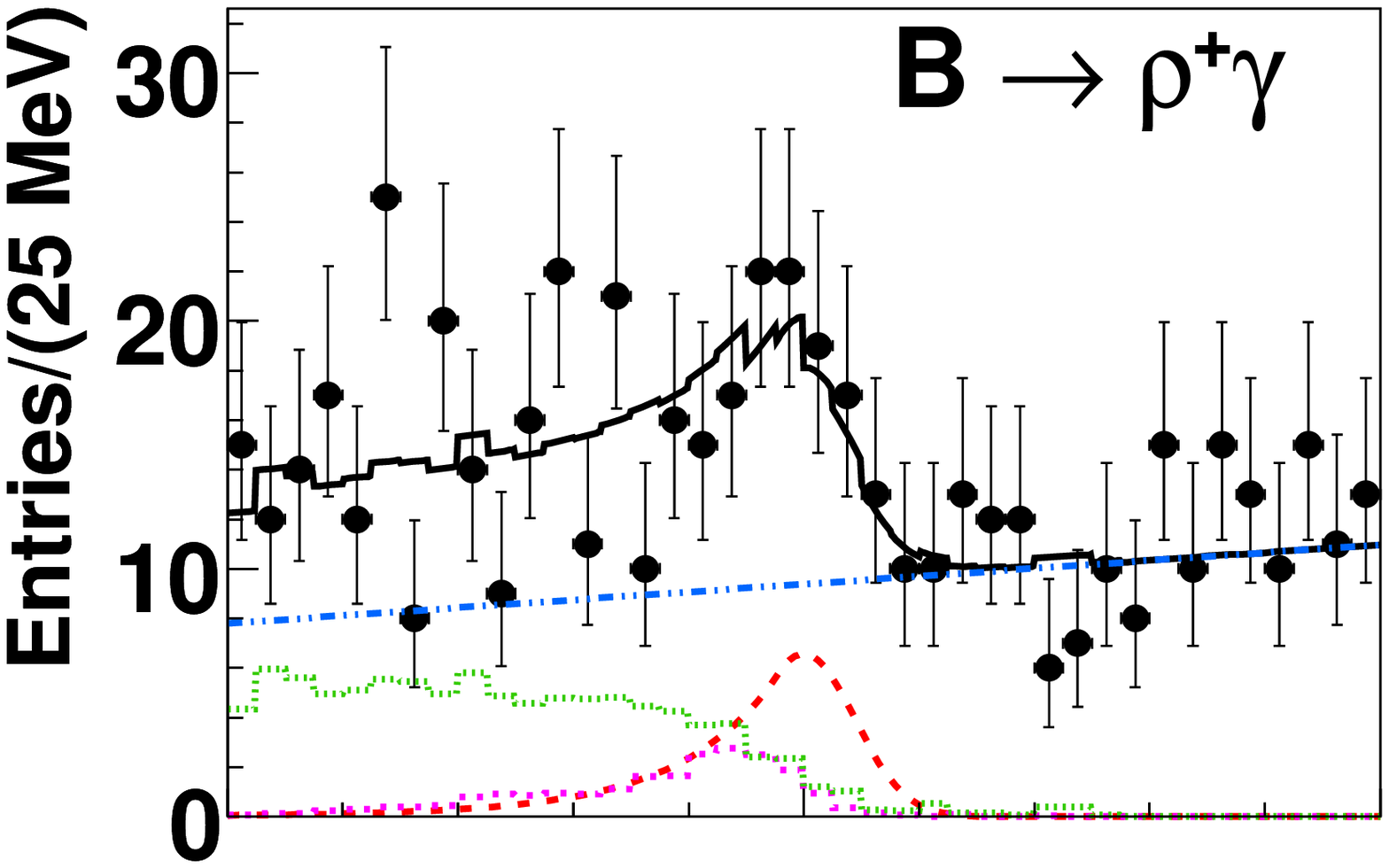}\\
       \myeps[\figtwoscale]{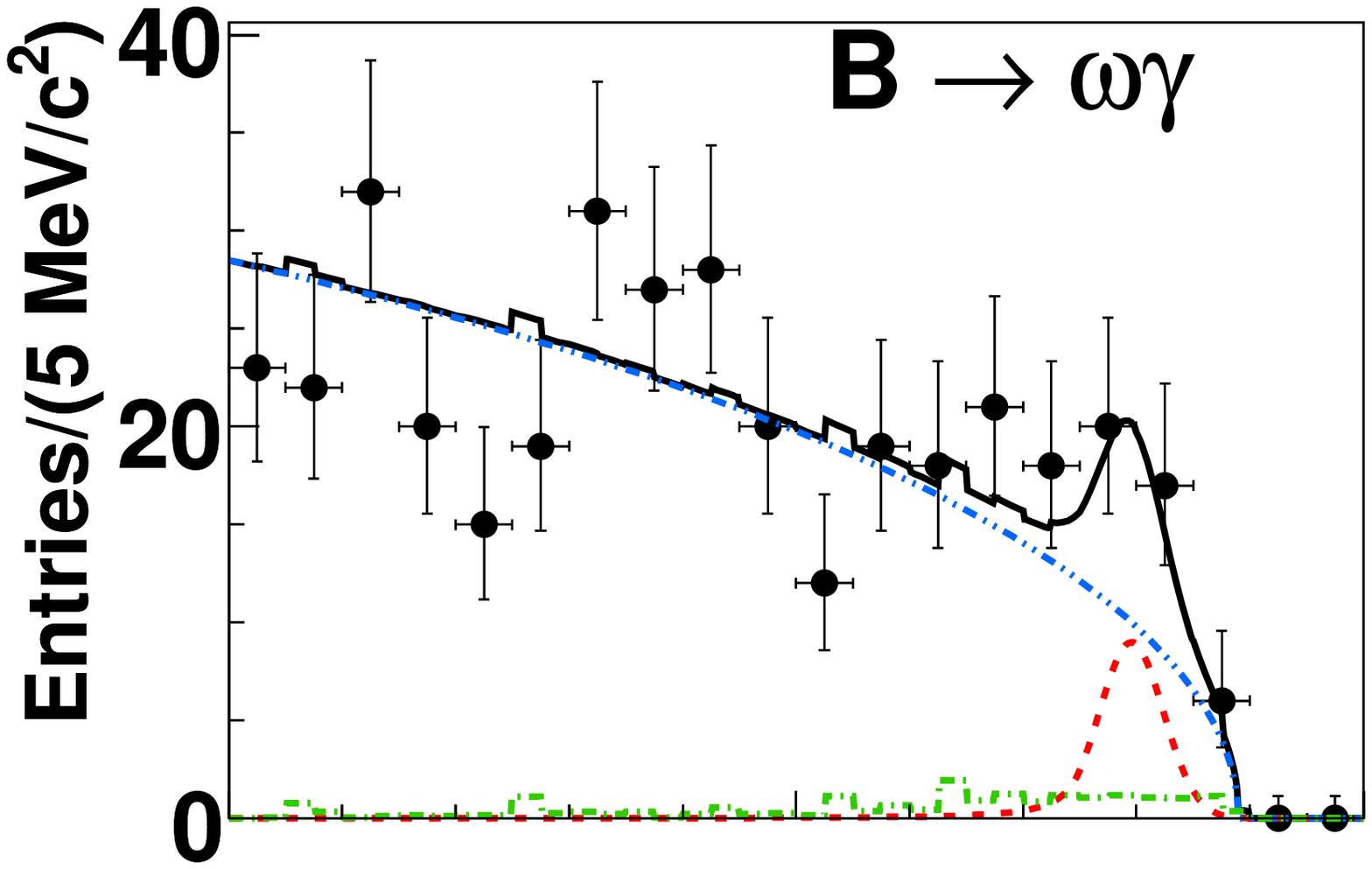}%
       \myeps[\figtwoscale]{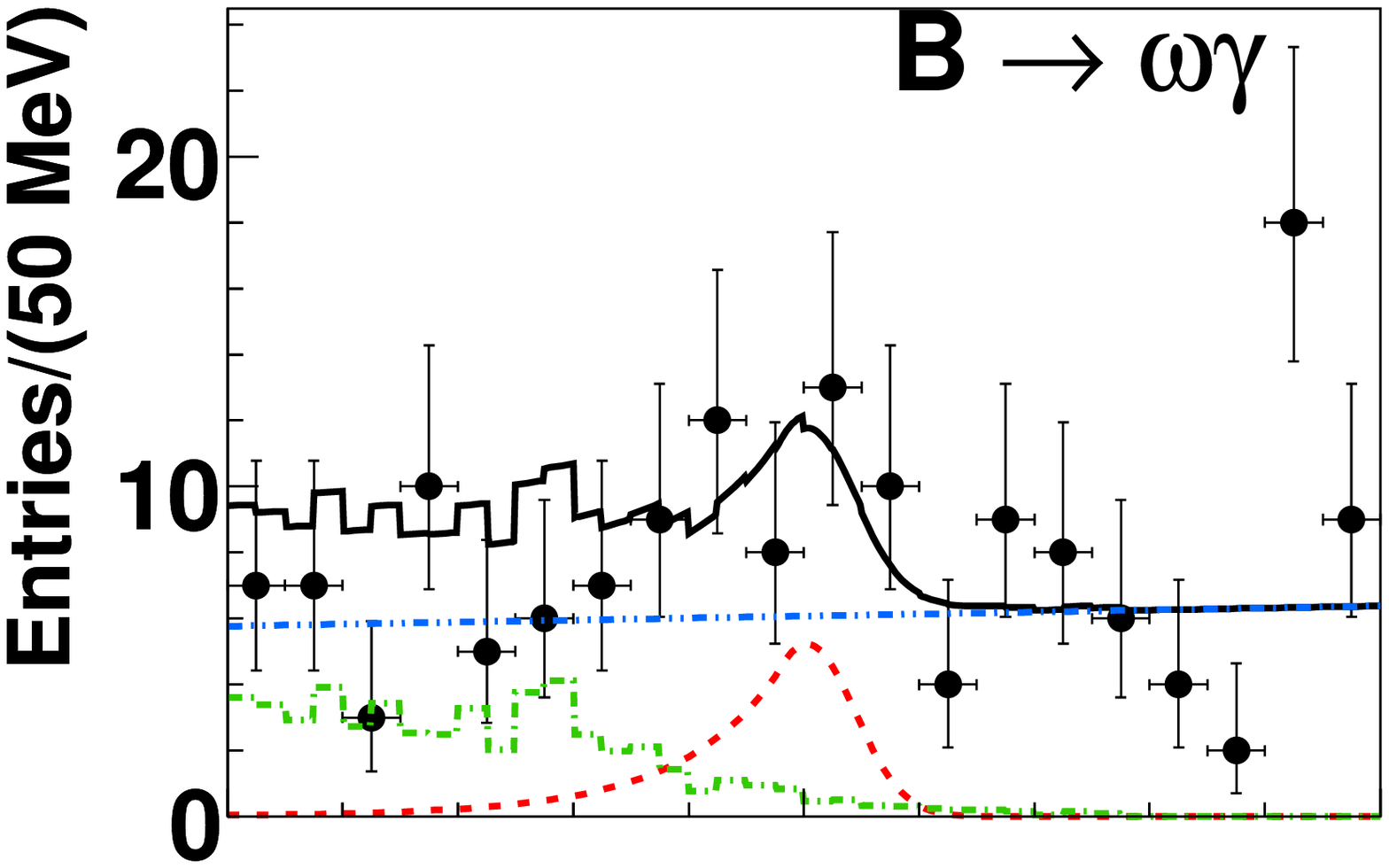}\\
       \myeps[\figtwoscale]{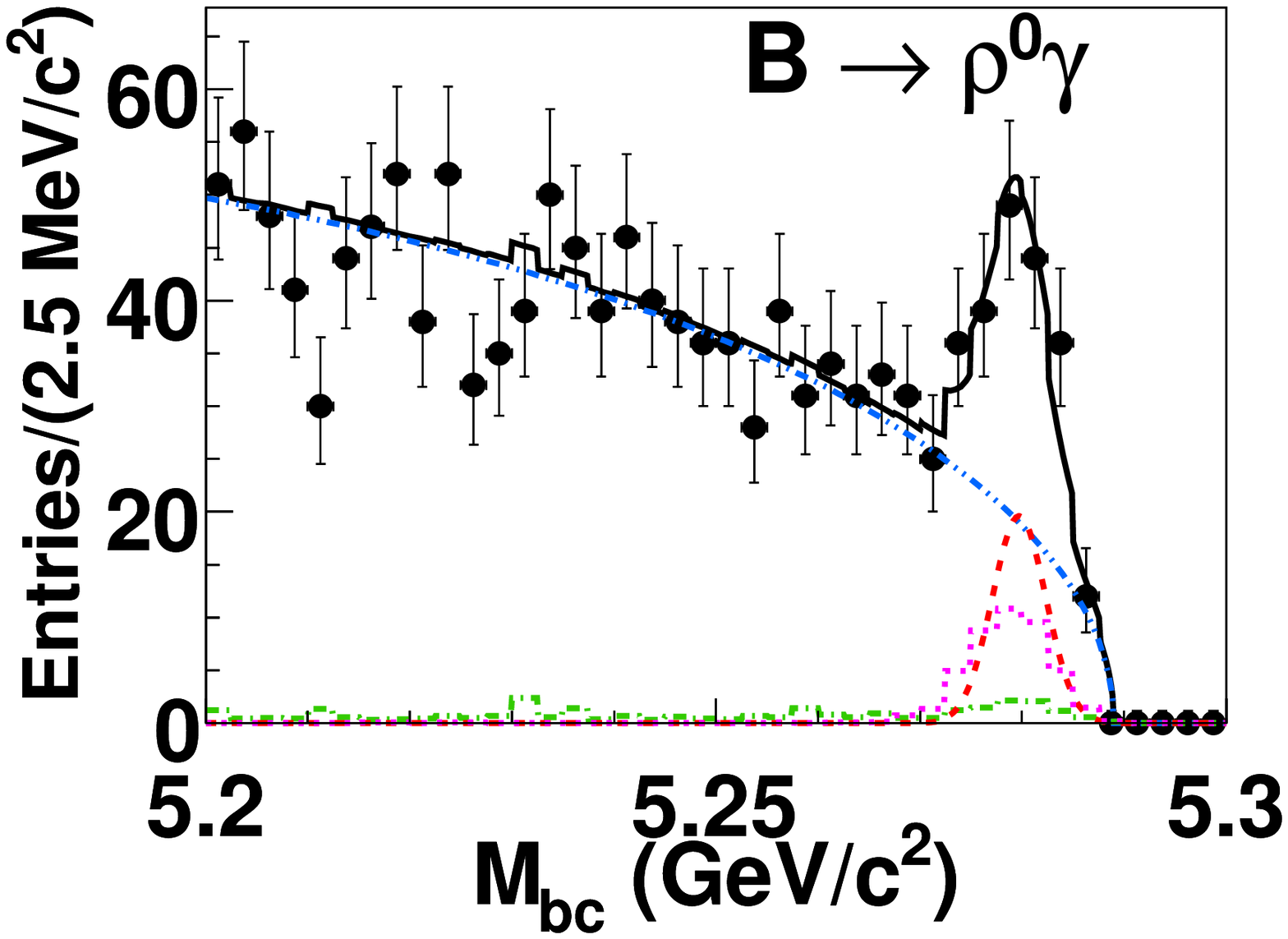}%
       \myeps[\figtwoscale]{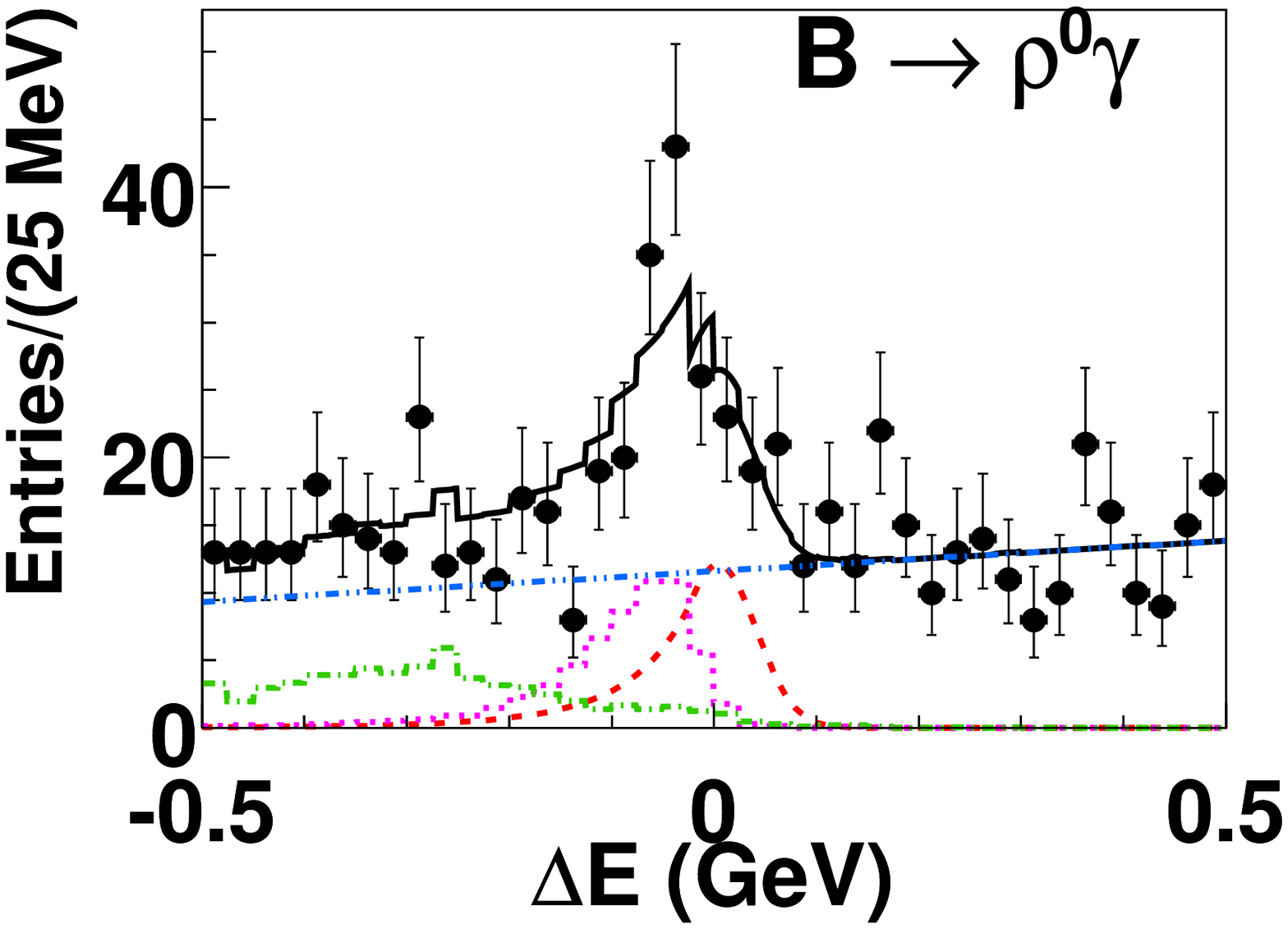} \\
        \vspace*{2pt}
       \myeps[\figtwoscale]{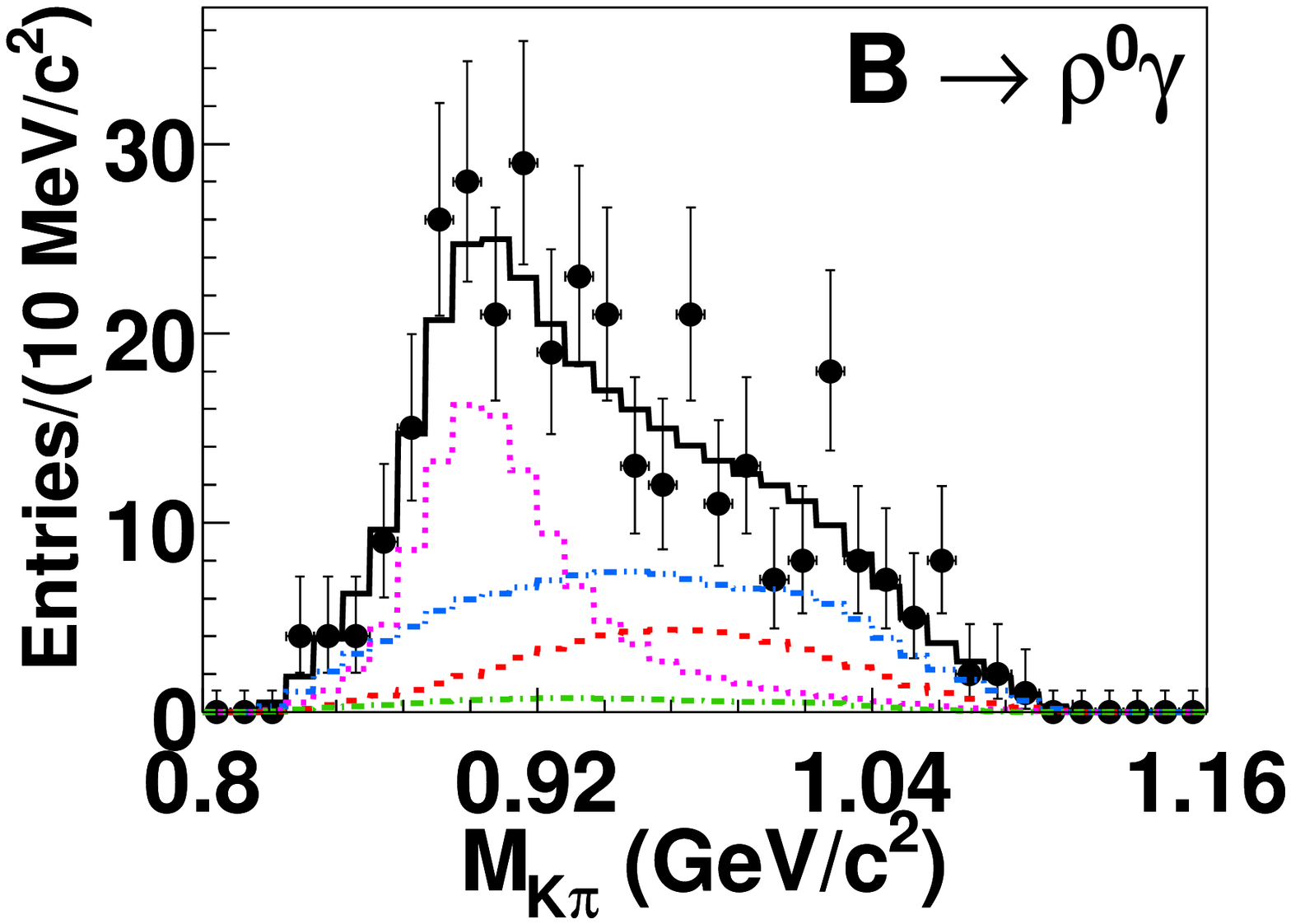}
     \vspace*{-5pt}
  \caption{Projections of the fit results to $\Mbc$ (in
  $|\DeltaE|<0.1\GeV$ and $0.92\GeVcc<\MKpi$), $\DeltaE$ (in
  $5.273\GeVcc<\Mbc<5.285\GeVcc$ and $0.92\GeVcc<\MKpi$), and for $\BtoRZG$,
  $\MKpi$.
  Curves show the signal (dashed, red), continuum (dot-dot-dashed, blue),
  $\BtoKG$ (dotted, magenta), other backgrounds (dash-dotted, green), and the total fit result
  (solid).}
  \label{fig:fit-rpom}
      \end{center}
   \end{figure}

The systematic error in the signal yield due to the fitting procedure
  is estimated by varying each of the fixed
parameters by $\pm1\sigma$ and then taking the quadratic sum of the
deviations in the branching fraction from the nominal value.  The varied
parameters are the signal shape parameters, branching fractions of the
background components, $\DeltaE$ shift of the $\KstarG$ component, and
the kaon to pion misidentification probability determined from a control
sample.
The results of the fits are shown in Fig.~\ref{fig:fit-rpom} and listed
in Table~\ref{tbl:results}.
\begin{table*}[ht]
\caption{Yield, significance with systematic
  uncertainty, efficiency, and branching fraction ($\Br$) for each
  mode.  The first and second errors in the yield and $\Br$ are
  statistical and systematic, respectively.  The sub-decay
 $\Br(\omega\to\pi^+\pi^-\pi^0)$ is included for the $\omega\gamma$ mode.
}
\label{tbl:results}
\begin{ruledtabular}
\begin{tabular}{lcccc}
Mode & Yield & Significance & Efficiency (\%) & $\Br$ ($10^{-7}$) \\
\hline
$\BtoRPG$ & $\NSBtoRPG$ & $\signifRPconv$ & $\EffRP$ & $\BrBtoRPG$ \\
$\BtoRZG$ & $\NSBtoRZG$ & $\signifRZconv$ & $\EffRZ$ & $\BrBtoRZG$ \\
$\BtoOMG$ & $\NSBtoOMG$ & $\signifOMconv$ & $\EffOM$ & $\BrBtoOMG$ \\
$\BtoRG$  & --- & $\signifRAconv$  & --- & $\BrBtoRAG$ \\
$\BtoROG$ & --- & $\signifROconv$  & --- & $\BrBtoROG$
 \\
\end{tabular}
\end{ruledtabular}
\end{table*}

The systematic error in the branching fraction has contributions from
the efficiency, fitting, and the number of $B$ meson pairs; we add
these together in  
quadrature.  The significance is defined as
$\sqrt{-2\ln(\Lzero/\Lmax)}$, where $\Lmax$ ($\Lzero$) is the value of
the likelihood function when the signal yield is floated (set to zero).
To include systematic uncertainty, the likelihood function from the fit is convolved with a Gaussian
systematic error function.

Table~\ref{tbl:results} also lists combined branching fractions, which
are calculated 
from the products of likelihoods from individual fits.  We combine
$\rhoPG$ and $\rhoZG$ modes (referred to as $\rhoG$) and three $\rhoG$
and $\omegaG$ modes (referred to as $\ROG$) 
 assuming a single branching fraction
$\Br(\BtoRG) \equiv
\Br(\BtoROG) \equiv \Br(\BtoRPG) = 2\times\tauBratio\Br(\BtoRZG) =
2\times\tauBratio\Br(\BtoOMG)$~\cite{bib:ali-1994,bib:ali-cdlu},
where $\tauBratio = 1.071\pm0.009$~\cite{bib:pdg2006}.
 The results are consistent with the
previous measurements~\cite{bib:belle-rhogam,bib:babar-rhogam} and have smaller
errors. They are also in agreement with SM
predictions~\cite{bib:ali-cdlu,bib:pball,bib:bosch-buchalla}.

The ratios of the branching fractions of the $B\to\rho\gamma/\omega\gamma$ modes to
those of the $B\to K^*\gamma$ modes can be related to
$|\Vtd/\Vts|$~\cite{bib:ali-cdlu,bib:pball}.
We calculate the ratios from likelihood curves of individual fits to
 the $B\to\rho\gamma/\omega\gamma$ and $B\to K^*\gamma$ samples.
Systematic errors that do not cancel in the ratio are convolved into the likelihoods.
We find,
\begin{eqnarray}
\frac{\Br(\BtoRZG)}{\Br(\BtoKZG)}&=&\ratioBtoRZGtoKZG,\\
\frac{\Br(\BtoRG)}{\Br(\BtoKG)}&=&\ratioBtoRGtoKG,\\
\frac{\Br(\BtoROG)}{\Br(\BtoKG)}&=&\ratioBtoROGtoKG,\label{eq:ratioROGKG}
\end{eqnarray}
\noindent
where the first and second errors are statistical and systematic,
respectively.

Using the prescription in Ref.~\cite{bib:pball}, Eq.~\ref{eq:ratioROGKG}
for example gives $|V_{td}/V_{ts}| = \VtdVtsrw$.
 This is consistent with determinations from  $B_s^0$
mixing ~\cite{bib:hfag2006},  which involve box diagrams rather
than penguin loops.
We also find $\Br(\BtoKPG) = (\BrBtoKPGstat)\EM7$ and $\Br(\BtoKZG) =
(\BrBtoKZGstat)\EM7$ (statistical error only), in agreement with 
the world average. 

From Table~\ref{tbl:results}, we calculate the isospin asymmetry
 $\Delta(\rho\gamma)=\tauBratiorev\Br(\BtoRPG)/\Br(\BtoRZG)-1$
 and find
\begin{eqnarray}
\Delta(\rho\gamma)&=&\AIBtoRG.
\end{eqnarray}
The result is in agreement with the previous
measurement~\cite{bib:babar-rhogam}, and is only marginally consistent
with the SM expectations~\cite{bib:pball,bib:alilungi}.

We also calculate 
 the direct $CP$-violating
asymmetry $\Acp(\BtoRPG)=[N(\rhoMG)-N(\rhoPG)]/[N(\rhoMG)+N(\rhoPG)]$
 using a simultaneous fit to $B^+\to\rho^+\gamma$ and
 $B^-\to\rho^-\gamma$ data samples. 
 We consider systematic errors due to
 the fitting procedure, asymmetries in the backgrounds,
  and possible detector bias estimated
 using a $B\to D\pi$ control sample.
 We use the measured asymmetries \cite{bib:hfag2006} for $B^+\to
 K^{*+}\gamma$, $\rho^+\pi^0$, $\rho^+\eta$ and $B\to X_s\gamma$ and   
  assume up to 100$\%$ asymmetry for other charmless hadronic
 $B$ decays.
 We find
\begin{eqnarray}
\Acp(\BtoRPG)&=&\AcpBtoRPG.
\end{eqnarray}
The result is consistent with the SM predictions~\cite{bib:pball,bib:ali-cdlu}.

In conclusion, we present a new measurement of branching fractions
for $\BtoRG$ and $\BtoOG$,  a measurement of the isospin asymmetry, 
 and the first measurement of the direct $CP$-violating asymmetry for $\BtoRPG$.
The results are consistent with SM predictions. 
 We improve the experimental precision on $|\Vtd/\Vts|$ determined
from penguin loops, finding good agreement with the value
determined from box diagrams~\cite{bib:hfag2006}.

We thank the KEKB group for excellent operation of the accelerator, the
KEK cryogenics group for efficient solenoid operations, and the KEK
computer group and the NII for valuable computing and SINET3
network support.  We acknowledge support from MEXT and JSPS (Japan); ARC
and DEST (Australia); NSFC and KIP of CAS (China); DST (India); MOEHRD,
KOSEF and KRF (Korea); KBN (Poland); MES and RFAAE (Russia); ARRS
(Slovenia); SNSF (Switzerland); NSC and MOE (Taiwan); and DOE (USA).


\begin{thebibliography}{99}
\bibitem{bib:rhogam-bsm}
For example,
A.~Arhrib, C.-K.~Chua and W.-S.~Hou, \Journal{\EPJC}{21}{567}{2001};
A.~Ali and E.~Lunghi, \Journal{\EPJC}{26}{195}{2002};
Z.-J.~Xiao and C.~Zhuang, \Journal{\EPJC}{33}{349}{2004}.

\bibitem{bib:belle-rhogam}
Belle Collaboration, D.~Mohapatra \etal, \Journal{\PRL}{96}{221601}{2006}.

\bibitem{bib:babar-rhogam}
Babar Collaboration, B.~Aubert \etal, \Journal{\PRL}{98}{151802}{2007}.

\bibitem{bib:ckm}
M. Kobayashi and T. Maskawa, \Journal{Prog. Theor. Phys.}{49}{652}{1973};
N. Cabibbo, \Journal{\PRL}{10}{531}{1963}.

\bibitem{bib:ckmfitter}
CKM fitter group, J.~Charles \etal, \Journal{\EPJC}{41}{1}{2005} and
updates at http://ckmfitter.in2p3.fr;
UTfit Collaboration, M. Bona \etal, arXiv:0707.0636 [hep-ph].

\bibitem{bib:pball}
P.~Ball, G.W.~Jones, and R.~Zwicky, J. High Energy Phys. 04 (2006) 046;
P.~Ball, G.W.~Jones, and R.~Zwicky,  \Journal{\PRD}{75}{054004}{2007}.

\bibitem{bib:alilungi}
A.~Ali and E.~Lunghi, \Journal{\EPJC}{26}{195}{2002}.

\bibitem{bib:kekb}
S.~Kurokawa and E.~Kikutani, \Journal{\NIMA}{499}{1}{2003}, and other
papers included in this Volume.

\bibitem{bib:belle-detector}
Belle Collaboration, A.~Abashian \etal, \Journal{\NIMA}{479}{117}{2002}.

\bibitem{bib:ksfw}
 Belle Collaboration, S.H.~Lee \etal, \Journal{\PRL}{91}{261801}{2003}.
        

\bibitem{bib:hamlet}
H.~Kakuno \etal, \Journal{\NIMA}{533}{516}{2004}.  


\bibitem{bib:cbls}
Crystal Ball Collaboration, J.~E.~Gaiser \etal, \Journal{\PRD}{34}{711}{1986}.

\bibitem{bib:argus-function}
ARGUS Collaboration, H.~Albrecht \etal, \Journal{\PLB}{241}{278}{1990}.

\bibitem{bib:hfag2006}
Heavy Flavor Averaging Group, winter 2006 results,
(http://www.slac.stanford.edu/xorg/hfag/).

\bibitem{bib:ali-1994} 
A.~Ali, V. M.~Braun and H.~Simma, \Journal{\ZPC}{63}{437}{1994}.

\bibitem{bib:ali-cdlu}
A.~Ali and A.~Parkhomenko,    \Journal{\EPJC}{23}{89}{2002};
C.-D.~Lu, M.~Matsumori, A.~I.~Sanda and M.-Z.~Yang,~
        \Journal{\PRD}{72}{094005}{2005}, Erratum-ibid D {\bf 73},~ 039902~ (2006).

\bibitem{bib:pdg2006}
W.-M.~Yao \etal, \Journal{\JPG}{33}{1}{2006}.

\bibitem{bib:bosch-buchalla}
In addition to Ref.~\cite{bib:ali-cdlu}, see e.g.,
S.~Bosch and G.~Buchalla, \Journal{\NPB}{621}{459}{2002};
T.~Huang, Z.~Li and H.~Zhang, \Journal{\JPG}{25}{1179}{1999};
R.~Fleischer and S.~Recksiegel, \Journal{\PRD}{71}{051501(R)}{2005};
M.~Beneke, T.~Feldmann and D.~Seidel, \Journal{\EPJC}{41}{173}{2005}.

\end{thebibliography}
\end{document}